\documentclass[aps, prl, twocolumn, superscriptaddress, 10pt, floatfix]{revtex4-2}
\usepackage{graphicx,float,amsmath,amssymb}
\usepackage[hidelinks]{hyperref}
\usepackage[scr=boondoxo,scrscaled=1.05]{mathalfa}
\usepackage{color}

\begin{document}

\setcitestyle{super}

\title{Heteroatomic Andreev molecule in a superconducting island-double quantum dot hybrid}

\author{Oliv\'er K\"urt\"ossy}
\affiliation{Department of Physics, Institute of Physics, Budapest University of Technology and Economics, M\H{u}egyetem rkp. 3, H-1111 Budapest, Hungary}

\author{Mih\'aly Bod\'ocs}
\affiliation{Department of Physics, Institute of Physics, Budapest University of Technology and Economics, M\H{u}egyetem rkp. 3, H-1111 Budapest, Hungary}

\affiliation{MTA-BME Nanoelectronics Momentum Research Group, M\H{u}egyetem rkp. 3, H-1111 Budapest, Hungary}

\author{Catalin Pascu Moca}
\affiliation{HUN-REN—BME Quantum Dynamics and Correlations Research Group,
Budapest University of Technology and Economics, M\H{u}egyetem rkp. 3, H-1111 Budapest, Hungary}

\affiliation{Department of Physics, University of Oradea, 410087, Oradea, Romania}

\author{Zolt\'an Scher\"ubl}
\affiliation{Department of Physics, Institute of Physics, Budapest University of Technology and Economics, M\H{u}egyetem rkp. 3, H-1111 Budapest, Hungary}

\affiliation{MTA-BME Nanoelectronics Momentum Research Group, M\H{u}egyetem rkp. 3, H-1111 Budapest, Hungary}

\author{Ella Nikodem}
\affiliation{Physics Institute II, University of Cologne, Zülpicher Str. 77, 50937 Cologne, Germany}

\author{Thomas Kanne}
\affiliation{Center for Quantum Devices, Niels Bohr Institute, University of Copenhagen, 2100 Copenhagen, Denmark}

\author{Jesper Nyg{\aa}rd}
\affiliation{Center for Quantum Devices, Niels Bohr Institute, University of  Copenhagen, 2100 Copenhagen, Denmark}

\author{Gergely Zar\'and}
\affiliation{HUN-REN—BME Quantum Dynamics and Correlations Research Group,
Budapest University of Technology and Economics, M\H{u}egyetem rkp. 3, H-1111 Budapest, Hungary}

\affiliation{Department of Theoretical Physics, Institute of Physics,
Budapest University of Technology and Economics, M\H{u}egyetem rkp. 3, H-1111 Budapest, Hungary}

\author{P\'eter Makk}
\email{makk.peter@ttk.bme.hu}
\affiliation{Department of Physics, Institute of Physics, Budapest University of Technology and Economics, M\H{u}egyetem rkp. 3, H-1111 Budapest, Hungary}

\affiliation{MTA-BME Correlated van der Waals Structures Momentum Research Group, M\H{u}egyetem rkp. 3, H-1111 Budapest, Hungary}

\author{Szabolcs Csonka}
\email{csonka.szabolcs@ttk.bme.hu}
\affiliation{Department of Physics, Institute of Physics, Budapest University of Technology and Economics, M\H{u}egyetem rkp. 3, H-1111 Budapest, Hungary}

\affiliation{MTA-BME Nanoelectronics Momentum Research Group, M\H{u}egyetem rkp. 3, H-1111 Budapest, Hungary}

\date{\today}

\pacs{}


\begin{abstract}
Topological superconductors (SCs) hold great promise for fault-tolerant quantum hardware, however, their experimental realization is very challenging. Recently, superconducting artificial molecules (Andreev molecules) have opened new avenues to engineer topological superconducting materials. In this work, we demonstrate a heteroatomic Andreev molecule, where two normal artificial atoms
realized by quantum dots (QDs) are coupled by a superconducting island (SCI). We show that the two normal atoms strongly hybridize and form a 3-electron-based molecular state. Our density matrix renormalization group (DMRG) calculations explain quantitatively the robust binding of electrons. The tunability of the structure allows us to drive a quantum phase transition from an antiferromagnetic Andreev molecular state to a heteroatomic Andreev molecule with ferromagnetically coupled QDs using simple electrical gating.
\end{abstract}

\maketitle

\section{Introduction}

Advancement in the realization of superconducting circuits granted the possibility to construct the first synthetic, so-called Andreev molecules, where two artificial states are coupled by an SC, similarly to conventional molecules formed by the hybridization of adjacent atoms. These superconducting molecules open new avenues for quantum hardware as they constitute the main operational units of topological quantum computing\cite{kitaev2003fault,nayak2008non} circuits based on non-abelian Majorana excitations\cite{lutchyn2010majorana,oreg2010helical,mourik2012signatures,das2012zero,deng2016majorana,albrecht2016exponential,thakurathi2018majorana,prada2020andreev,zatelli2023robust}. 

When an SC electrode is coupled to a normal conductor or an artificial atom, the superconducting correlations leak into them and Yu-Shiba Rusinov (YSR) or Andreev states form\cite{yu1965bound,shiba1968classical,rusinov1969theory,buitelaar2002quantum,balatsky2006impurity,sand2007kondo,eichler2007even,grove2009superconductivity,lee2014spin,jellinggaard2016tuning,scherubl2020large}. Recent experiments demonstrated the crossed Andreev reflection induced hybridization of Josephson junctions\cite{haxell2023demonstration,matsuo2023phasedep,matsuo2023phase}, level-tunable artificial atoms\cite{kurtossy2021andreev}, different Andreev\cite{coraiola2023phase,junger2023intermediate} and Yu-Shiba-Rusinov (YSR) dimers\cite{kezilebieke2018coupled,ruby2018wave,choi2018influence,beck2021spin,ding2021tuning}. The common concept in these works is that a bulk SC, playing the role of a Cooper pair reservoir, mediates the interaction between two normal regions\cite{flatte2000local,lesovik2001electronic,recher2002superconductor},
and the structure of the QD-SC-QD system resembles that of a H\textsubscript{2} molecule.

\begin{figure}[h!]
\begin{center}
\includegraphics[width=0.5\textwidth]{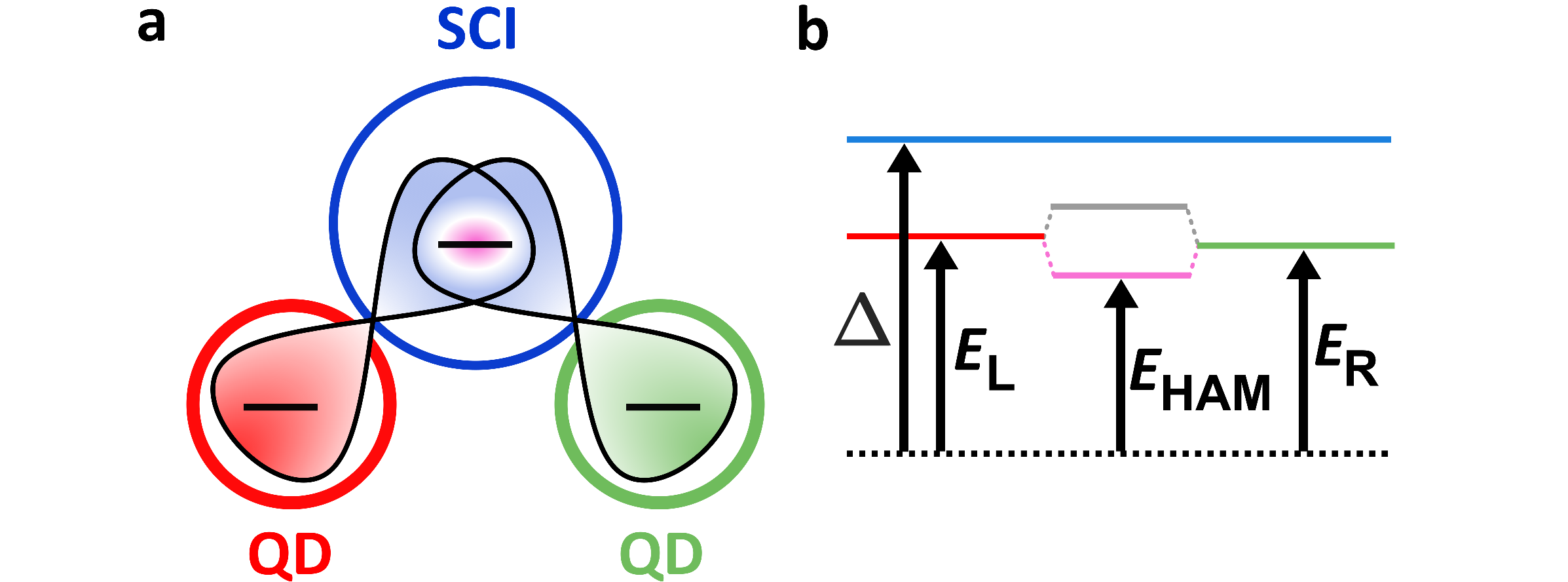}
\caption{\label{multi_island_shiba_molecule_schematics} \textbf{Energy schemes of different YSR states.} \textbf{a} Concept of a heteroatomic Andreev molecule. The levels of two QDs (red, green) couple to the SCI 
and hybridize (pink), thereby forming a 3-particle state. \textbf{b} Energy scheme of a coupled SCI-double QD system. The hybridization of the red and green YSR states at energies $E_\mathrm{L}$ and $E_\mathrm{R}$ results in bonding and anti-bonding mixed states splitting in energy. The lower one becomes the heteroatomic Andreev molecule at $E_\mathrm{HAM}$ (pink). The superconducting gap is labeled by $\Delta$ (blue).}
\end{center}
\end{figure}

The picture changes qualitatively if the size of the SC is finite. Coulomb repulsion becomes significant with scaling down the dimensions, yielding a superconducting island (SCI), where single electron charging and pair correlations compete\cite{tuominen1992experimental,averin1992single,eiles1993even,lafarge1993measurement,joyez1994observation}. As a result, the SCI can have an unpaired electron in stark contrast to bulk superconductors. Recently it has been shown that a single quasiparticle of an SCI can bind to an impurity establishing a Coulomb-aided YSR singlet\cite{pavevsic2021subgap,estrada2022excitations,estrada2024correlation}. Exploiting this exchange-like interaction, one can think of a novel approach of coupling two QDs via an SCI, which acts as a distinct, central atom as introduced in Fig. \ref{multi_island_shiba_molecule_schematics}\textbf{a}. Here the screening quasiparticle of the SCI is shared between two YSR states forming a 3-body state in a peculiar way, which we call heteroatomic Andreev molecule. This bound state can exist at energy $E_\mathrm{HAM}$, lying lower than both single YSR states ($E_\mathrm{L(R)}$) and the superconducting gap, $\Delta$, as sketched in Fig. \ref{multi_island_shiba_molecule_schematics}\textbf{b}. In the language of molecular physics, this structure resembles the H\textsubscript{2}O molecule.

In this paper, we demonstrate the experimental signature of a heteroatomic Andreev molecule hosted by an SCI-double QD hybrid realized in parallel InAs nanowires. We utilize the Coulomb blockade spectroscopy as a tool to capture the excitation energies of different electron configurations in the SCI and the QDs confirming the presence of a 3-electron hybrid state. The gate tunability of the SCI allows us to drive a quantum phase transition between 2-body Andreev states and 
3-electron heteroatomic Andreev molecular states. The main experimental findings are reproduced by simple numerical simulations, as well as by DMRG calculations. Moreover, our model reveals the different spin configurations of the heteroatomic Andreev molecule in terms of exchange interaction, which can be changed from antiferromagnetic to ferromagnetic as an unpaired quasiparticle is added to the SCI. The results show that this novel H\textsubscript{2}O architecture can be robustly realized in artificial quantum circuits and polymerization of the SCI-QD system can be used to construct longer chains for topological quantum circuits.

\begin{figure*}[htp]
\begin{center}
\includegraphics[width=1\textwidth]{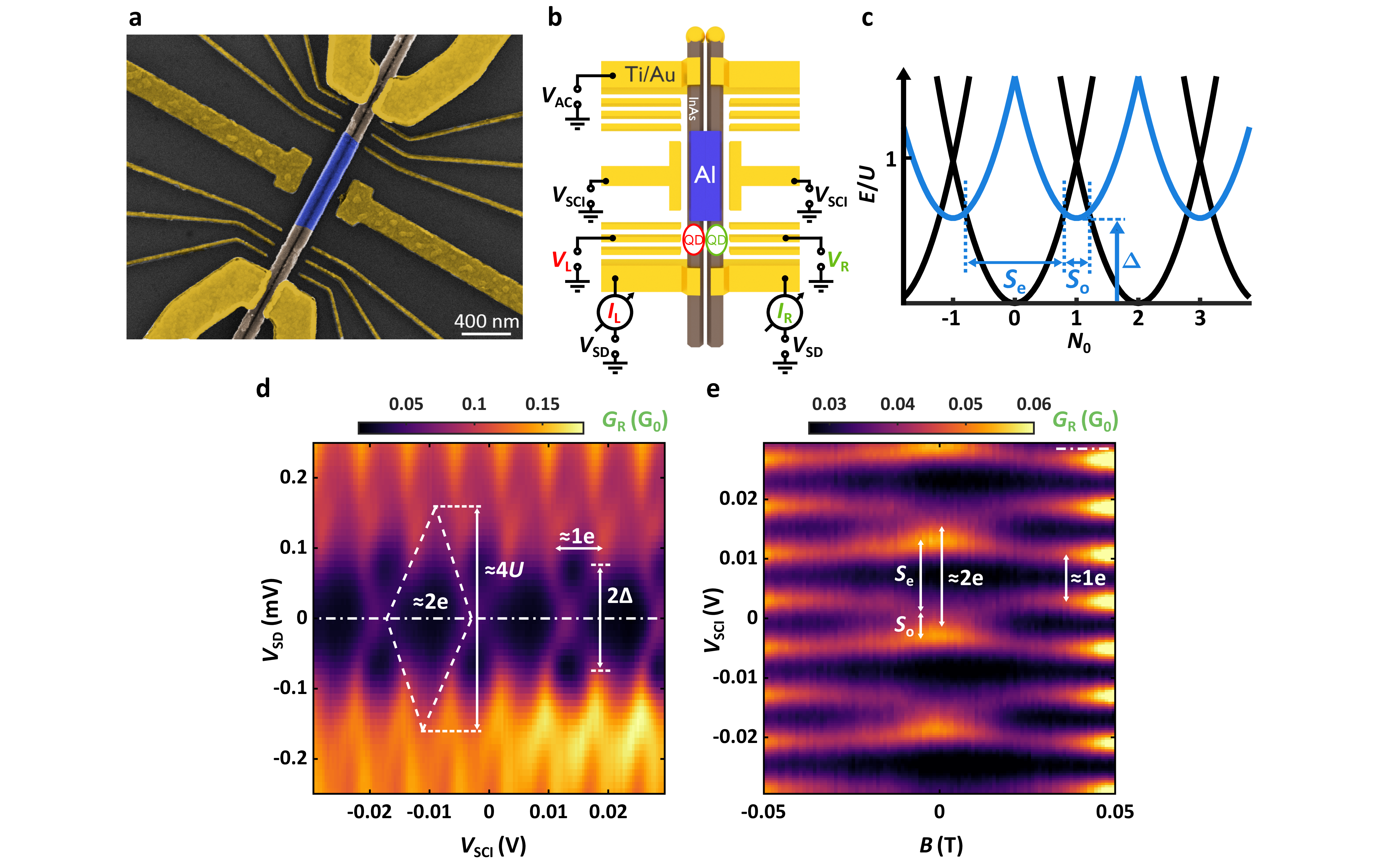}
\caption{\label{multi_island_device_and_island} \textbf{Device outline \& SCI characteristic.} \textbf{a} SEM micrograph of the device measured in multiple terminals. The epitaxial Al (blue) was etched away along the wires except in the middle, thereby forming an island connecting the separate InAs nanowires (brown). Four Ti/Au electrodes were installed as normal contacts and finger gates to gain a high level transport control. \textbf{b} Schematic illustration of the measurement setup. The AC source was applied to the top left contact, while the differential conductance was measured simultaneously on the bottom left (red) and bottom right (green) drain electrodes. Level positions of the red and green QDs were tuned by plunger gate voltages $V_\mathrm{L}$ and $V_\mathrm{R}$, respectively. The SCI was gated by $V_\mathrm{SCI}$. \textbf{c} Energy diagram of a decoupled SCI island. Blue parabolas shifted up by $\Delta$ correspond to odd parity states with ground state spacing $S_o$. \textbf{d} Coulomb blockade spectroscopy of the SCI through the bottom right drain. $\Delta\approx U$ yields an intrinsic close-to-2e periodicty with diamonds of height $\approx 4U$. $2\Delta$ is determined from the onset of the 1e-periodic patterns in the energy spectrum. \textbf{e} Evolution of zero-bias SCI resonances (along the white dash-dotted line of panel \textbf{d}) in an out-of-plane magnetic field. The 2e-periodic pattern gradually turns into 1e-periodicity as superconductivity is destroyed.}
\end{center}
\end{figure*}

\section{Device outline}

The investigated system is shown in Figs. \ref{multi_island_device_and_island}\textbf{a-b}. A pair of parallel InAs nanowires (brown) were connected by a $\approx 700\,$nm long SCI as shown in the scanning electron micrograph (SEM) in Fig. \ref{multi_island_device_and_island}\textbf{a}. Four Ti/Au electrodes (yellow) were defined such that each one contacted only one nanowire segment individually, while finger gates were installed surrounding the nanowires to confine the QDs. Electronic transport measurements were performed at a base temperature of $40\,$mK (for details, see Methods) with QDs formed in the bottom left (red, labeled by "L") and bottom right (green, "R") segments as illustrated in Fig. \ref{multi_island_device_and_island}\textbf{b}. The top left electrode was biased with $V_\mathrm{AC}$ as a source, the top right was floated, and the rest acted as drains biased with DC voltage $V_\mathrm{SD}$. Differential conductances $G_\mathrm{R}$ and $G_\mathrm{L}$ in the bottom left and right branches were measured simultaneously via the red and green QDs, respectively. In this setup, effectively 2 parallel channels were probed: one of them consisted of the SCI and the red QD, the other one the SCI and the green QD in series.

If an SCI is decoupled from the environment, the electron number on it ($N_0$) becomes quantized as in a regular QD, but the energy dispersion is characterized by the ratio of the superconducting gap, $\Delta$, and the charging energy, $U$. For $\Delta >U$, the ground state has an even number of electrons at any gate voltage, and the SCI's energy follows the black parabolas in Fig. \ref{multi_island_device_and_island}\textbf{c}. However, for $\Delta <U$, odd occupations with one unpaired quasiparticle of energy $\Delta$ are also allowed, 
yielding the blue lines intersecting the black parabolas in Fig. \ref{multi_island_device_and_island}\textbf{c}.\cite{tuominen1992experimental,averin1992single,eiles1993even,lafarge1993measurement,joyez1994observation}. 
Consequently, the size $S_{e/o}$ of even/odd Coulomb diamonds alternates with
with the ground state parity\cite{lafarge1993measurement} ($S_o$ for odd and $S_e$ for even) 
\cite{lafarge1993measurement}
\begin{equation}\label{eq:spacing}
\frac{S_o}{S_e}=\frac{U-\Delta}{U+\Delta},
\end{equation} 
referred to as the even-odd effect. We remark that if a sub-gap state exists below the SCI, it governs the lowest-lying excitation at energy $E_0$ instead of $\Delta$ as reported in previous works\cite{higginbotham2015parity,albrecht2016exponential,albrecht2017transport,shen2018parity,o2018hybridization,vekris2022electronic}.

To characterize our SCI and to determine $\Delta$ and $U$, we accomplished finite-bias spectroscopy as a function of plunger gate voltage $V_\mathrm{SCI}$ through the bottom right arm, shown in Fig. \ref{multi_island_device_and_island}\textbf{d}. Here the QDs were decoupled from the islands and were set deep in Coulomb blockade to serve as co-tunneling probes. Within the white dashed lines, $N_0$ is even, while the odd states can not be resolved suggesting the close-to-2e periodic limit in the SCI diamonds with $4U$ total height\cite{hergenrother1994charge}. The lowest bias voltage where 1e periodic pattern appears is assigned to an excitation energy $\Delta$. We estimate
$U=85\,\mu$eV and $\Delta =75\,\mu$eV from the spectrum. 
Applying an out-of-plane magnetic field suppresses the superconductivity, hence the even-odd effect vanishes continuously, as presented in Fig. \ref{multi_island_device_and_island}\textbf{e}; the 2e-periodic signal at $B=0$ develops first at intermediate fields into even-odd oscillations with spacings $S_{o/e}$ for the odd/even states, and turns into a 1e-periodic signal at large fields, typical for normal metallic islands\cite{tuominen1992experimental,averin1992single,eiles1993even,lafarge1993measurement}.

\section{Results}

Now we explore the interaction between the SCI and the 2 QDs by coupling them strongly. We recorded the zero-bias conductance of the SCI and the green (red) QD controlled by their plunger gate voltages, $V_\mathrm{SCI}$ and $V_\mathrm{R(L)}$, while fixing the on-site energy of the red (green) QD. 
This reduces the problem to a double-QD stability diagram whose structure can be examined as a function of the occupation of the 3rd (untuned) QD (the characterization of the QDs can be found in Supplementary Note 1). A further advantage of this routine is that the even-odd amplitude $S_o/S_e$ of the SCI (see Eq. \ref{eq:spacing}) can be directly extracted from Coulomb-blockade spectroscopy for a given QD configuration, which reflects the energy cost of adding an unpaired electron to the SCI according to Eq. \ref{eq:spacing}.

\begin{figure*}[htp]
\begin{center}
\includegraphics[width=1\textwidth]{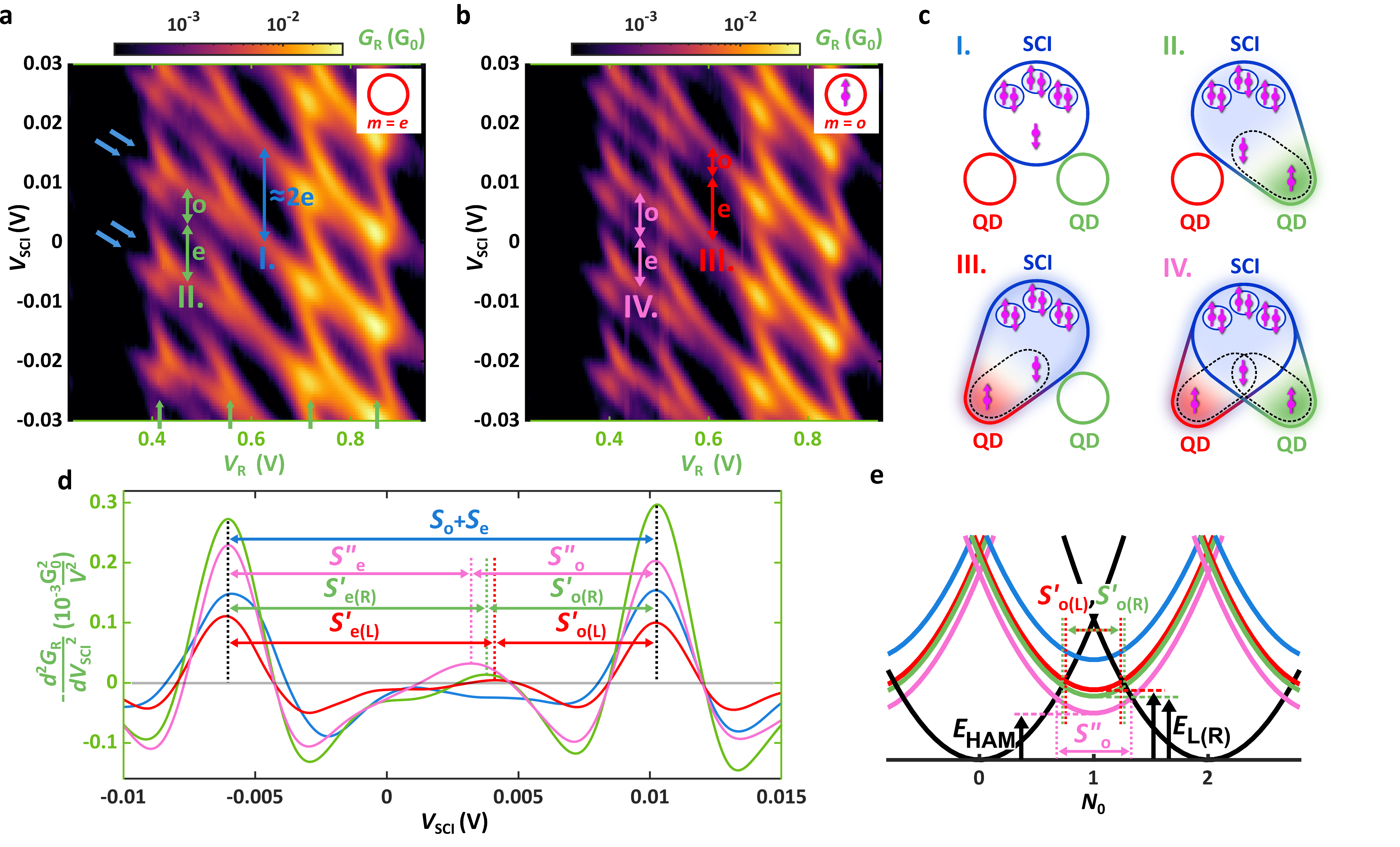}
\caption{\label{multi_island_measurement_br} \textbf{Stability diagrams exploring the Shiba molecule.} \textbf{a} Zero-bias stability map vs $V_\mathrm{R}$ and $V_\mathrm{SCI}$ via the SCI-green double QD with even number of electrons in the red QD. For $\vert e,N_0,e\rangle$ ($V_\mathrm{R}=0.6\,$V), nearly 2e charging, for $\vert e,N_0,o\rangle$, even-odd effect is obtained on the SCI. \textbf{b} Same as \textbf{a}, but captured at odd occupation of the red QD. The $\vert o,N_0,e\rangle$ state at $V_\mathrm{R}=0.6\,$V exhibits roughly the same diamond spacing as $\vert e,N_0,o\rangle$, the odd state of the SCI is extended at $V_\mathrm{R}=0.45\,$V when the filling of the QDs and the SCI is $\vert o,N_0,o\rangle$. \textbf{c} Illustration of the interaction between the QDs and a single quasiparticle in the SCI. While scenario \textbf{I.} with $\vert e,o,e\rangle$ represents a non-interacting picture, scenarios \textbf{II.} and \textbf{III.} with $\vert e,o,o\rangle$ and $\vert o,o,e\rangle$, yield distinct YSR in the red and green QDs. In the case of \textbf{IV.} with $\vert o,o,o\rangle$, the two YSR states share the unpaired electron. \textbf{d} Peak analysis of SCI resonance lines taken along the colored arrows in panels \textbf{a-b}. $S_o''>S_{o\mathrm{(L)[R]}}'$ fulfills the expectation predicting a heteroatomic Andreev molecule. \textbf{e} Energy diagram of the YSR states and the heteroatomic Andreev molecule. The latter one is set in deeper energy thereby shifting the pink parabola down by $E_\mathrm{R}-E_\mathrm{HAM}$ and broadening the odd state in the SCI to $S_o''>S_{o\mathrm{(L)[R]}}'$.}
\end{center}
\end{figure*}

The conductance $G_\mathrm{R}$, presented in Fig. \ref{multi_island_measurement_br}\textbf{a} as a function of $V_\mathrm{R}$ and $V_\mathrm{SCI}$, exhibits a characteristic honeycomb pattern, well-known for double QDs\cite{van2002electron}. The vertical resonance lines 
at $V_\mathrm{R}\approx cst$, indicated by the green arrows, are associated with the green QD's charge degeneracies, whereas the diagonal lines correspond to SCI charge degeneracies (blue arrows). Let us introduce the notation $\vert m,N_0,n\rangle =\vert m \rangle_\mathrm{L} \otimes \vert N_0 \rangle_\mathrm{SCI} \otimes \vert n \rangle_\mathrm{R} $, where $m, N_0, n=\lbrace e,o\rbrace$ express the parity of electron numbers in the red QD, the SCI, and the green QD with $e$ and $o$ addressing the even and odd occupations, respectively. In this particular measurement, the left (red) QD was set into blockade with an even number of electrons, thus $\vert e,N_0,n\rangle$ states were studied as indicated by the inset. At $V_\mathrm{R}\approx 0.6\,$V with $\vert e,N_0,e\rangle$, even number of electrons in both QDs, the SCI shows a close-to-2e charging behavior with spacing $S_e$ marked by the blue arrow (\textbf{I.}), similarly to Fig. \ref{multi_island_device_and_island}\textbf{c}. No YSR states are formed as illustrated in Fig. \ref{multi_island_measurement_br}\textbf{c} \textbf{I}. However, tuning the green QD to odd occupation at $V_\mathrm{R}\approx 0.45\,$V ($\vert e,N_0,o\rangle$ states), the resonance of the SCI changes drastically, it splits, and an even-odd effect is observable with a non-zero $S_{o\mathrm{(R)}}'$ and $S_{e\mathrm{(R)}}'<S_e$ spacings indicated by the green arrows (\textbf{II.}) in Fig. \ref{multi_island_measurement_br}\textbf{a}. The effective $\Delta$ in Eq. \ref{eq:spacing} is reduced to $E_\mathrm{R}$, suggesting the presence of a Coulomb-aided YSR singlet, composed by a quasiparticle in the SCI and the electron of the green QD ($\vert e,o,o\rangle$ state), as outlined 
in Ref. \citenum{estrada2022excitations}. 
Whereas the SCI and the green QD are strongly hybridized, the red QD does not interact with them, as sketched in panel 
Fig. \ref{multi_island_measurement_br}\textbf{c} \textbf{II.}.

We now examine how the stability diagram deviates if the red QD is filled with a single electron as well. Fig. \ref{multi_island_measurement_br}\textbf{b} demonstrates the same map as in panel \textbf{a} but recorded with $\vert o,N_0,n\rangle$ configurations. At $V_\mathrm{R}\approx 0.6\,$V, the close-to-2e charging observed in panel \textbf{a} 
is replaced with an even-odd pattern with $S_{o\mathrm{(L)}}'$ and $S_{e\mathrm{(L)}}'$, highlighted by the red arrows (\textbf{III.}). The even-odd amplitude observed is similar to the one characterizing the $\vert e,N_0,o\rangle$ state, \textbf{II.}. We conclude that, in this region, a different YSR state of character $\vert o,o,e\rangle$ and energy $E_\mathrm{L}\gtrapprox E_\mathrm{R}$ is formed between the SCI and the red QD (see Fig. \ref{multi_island_measurement_br}\textbf{c} \textbf{III.}).

Bringing both QDs to odd occupations, $\vert o,N_0,o\rangle$, visible at $V_\mathrm{R}\approx 0.45\,$V in panel \ref{multi_island_measurement_br}.\textbf{b}), the size $S_o''$ of the SCI odd state, $\vert o,o,o\rangle$, expands further, as revealed by the pink arrows (\textbf{IV.}). The stabilization of $S_o''$ entails excitation energy 
below both $E_\mathrm{L}$ and $E_\mathrm{R}$, and confirms the coupling of the green and red YSR states as shown by panel \textbf{c} \textbf{IV}. The same tendency was captured in both $G_\mathrm{L}$ and stability sweeps, with 
the role of the red and green QDs exchanged (see details in Supplementary Note 2).

\begin{figure*}[htp]
\begin{center}
\includegraphics[width=1\textwidth]{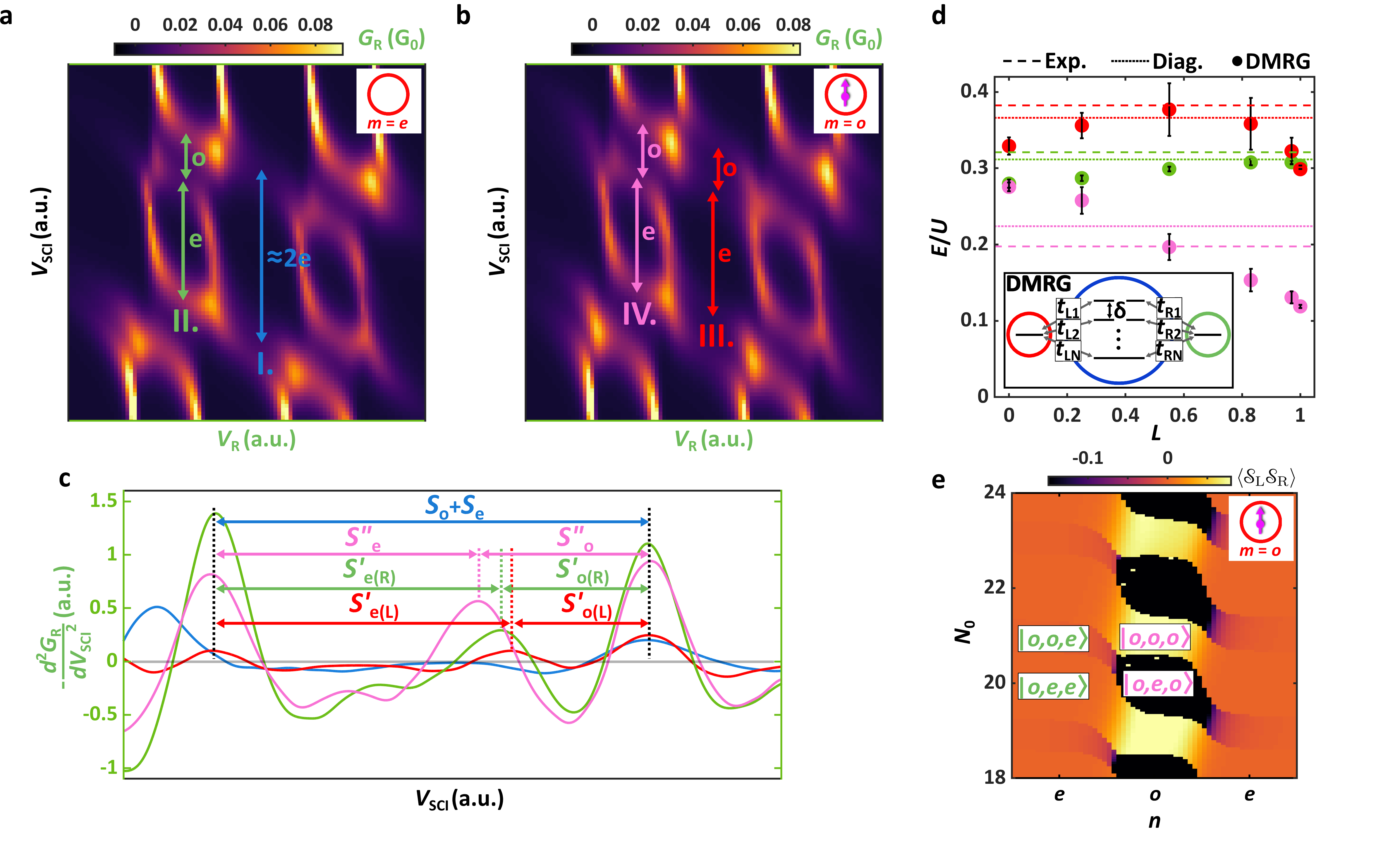}
\caption{\label{multi_island_simulation} \textbf{Simulation of a heteroatomic Andreev molecule.} \textbf{a} Calculated stability diagram of $G_\mathrm{R}$ versus $V_\mathrm{R}$ and $V_\mathrm{SCI}$ with the red QD being empty. The map mimics Fig. \ref{multi_island_measurement_br}\textbf{a}. \textbf{b} Same as panel \textbf{a}, but with the red QD filled with a single electron following the qualitative behavior in Fig. \ref{multi_island_measurement_br}\textbf{b}. \textbf{c} Peak analysis of the SCI resonances taken along the colored arrows in panels \textbf{a-b} matching the experimental data from Fig. \ref{multi_island_measurement_br}\textbf{d}. \textbf{d} Energy scheme of the Coulomb-aided YSR states and the heteroatomic Andreev molecule state versus $L$, as derived from DMRG calculations (solid lines). The red, green, and pink graphs belong to $E_\mathrm{L}$, $E_\mathrm{R}$, and $E_\mathrm{HAM}$, respectively. The experimental values and the energies extracted from the minimal model are depicted by the dashed and dotted lines matching DMRG at $L\approx0.55$. The inset illustrates the coupling mechanism between the QDs and the SCI orbitals. \textbf{e} Ground state spin correlation of the QDs (DMRG calculation) as a function of the QD and SCI electron configurations. The red QD is filled with a single electron ($m=o$). For $n=o$, the ordering is antiferromagnetic for the $\vert o,e,o\rangle$ states, while it is ferromagnetic for $\vert o,o,o\rangle$ configurations}.
\end{center}
\end{figure*}

To visualize the effect presented in Figs. \ref{multi_island_measurement_br}\textbf{a-c} and to obtain the distance of the SCI peak positions, $S_e$, $S_o$, $S_{o\mathrm{(L)[R]}}'$ and $S_o''$ precisely, we plot in Fig.\ref{multi_island_measurement_br}\textbf{d} the curvature
$p=-d^2G_\mathrm{R}/dV_\mathrm{SCI}^2$ along the colored arrows in \textbf{I.}-\textbf{IV.} for all 4 distinct QD parities, $m,n=\lbrace o,e\rbrace$.
Each curve of a certain color belongs to cuts taken along the arrow with the corresponding color. In the analysis, we consider only peaks with $p\geq 0$, corresponding to Coulomb blockade resonance positions. The $N_0=e\leftrightarrow o$ (secondary peaks) and the $N_0=o\leftrightarrow e$ transitions (main peaks) on the SCI are strongly asymmetric in amplitude. For transparency, we therefore center the main peaks of the curves at $V_\mathrm{SCI}=0.01\,$V and compare them accordingly. 

Manifestly, the blue line ($\vert e,N_0,e\rangle$ states) retains only the main peaks within our experimental resolution with $S_e+S_o=16.2\,$mV, corresponding to a close-to-2e charging. In the electrostatic picture, this state belongs to the blue parabola in Figs. \ref{multi_island_device_and_island}\textbf{c} and \ref{multi_island_measurement_br}\textbf{e}. In the red and green curves in panel \textbf{d} ($\vert o,N_0,e\rangle$ and $\vert e,N_0,o\rangle$ states) the secondary peaks appear at slightly different $V_\mathrm{SCI}$ values, providing $S_{o\mathrm{(L)}}'=5\,$mV and $S_{o\mathrm{(R)}}'=5.5\,$mV odd state widths. Acoording to Eq. \ref{eq:spacing}, the single Coulomb-aided YSR states of $\vert o,o,e\rangle$ and $\vert e,o,o\rangle$ reside in the QDs at energies $E_\mathrm{L}\approx32\,\mu$eV and $E_\mathrm{R}\approx 27\,\mu$eV, respectively. The energy of these states evolves along the red and green parabolas in panel \textbf{e}. The secondary peaks along the pink line of in panel \textbf{d} ($\vert o,N_0,o\rangle$ state) have substantially larger stability regions, $S_o''=6.5\,$mV, implying an even lower energy, $E_\mathrm{HAM}\approx 17\,\mu$eV. We interpret this increased binding energy as a result of the hybridization sketched in Fig. \ref{multi_island_shiba_molecule_schematics}\textbf{b}, pushing down the energies to the pink curve in panel \textbf{e}. Using 1e-periodicity as a reference, the relative deviation in the spacing of the two, coupled YSR states (case \textbf{IV.}) compared to the single one at lower energy (case \textbf{II.}) is significant, $\Delta S=2(S_o''-S_{o\mathrm{(R)}}')/(S_e+S_o)\approx 12$\%, 
and is consistent with the formation of a heteroatomic Andreev molecule in the $\vert o,o,o\rangle$ configuration.

\section{Discussion}

To confirm the existence of the heteroatomic state, we developed a simple QD-SC-QD model ("mixed orbital Andreson model) to reproduce the main experimental findings. In our calculations, the red and green QDs were represented by single and a two-orbital Anderson models. 
The SCI was described at the level of a two-orbital Richardson Hamiltonian, 
tunnel-coupled to both QDs. The eigenstates of the system were derived by exact diagonalization, whereas the transport was computed
by a simple rate equation model, assuming normal electrodes 
coupled to the SCI and the QDs. Details of the model can be found in Supplementary Note 3.

Figs. 4\textbf{a-b} show the simulated stability diagram of the system, replicating Figs. \ref{multi_island_measurement_br}\textbf{a-b} in a narrower range of $V_\mathrm{SCI}$. The close-to-2e charging of the $\vert e,N_0,e\rangle$ states (blue, \textbf{I.} from Fig. \ref{multi_island_measurement_br}\textbf{c}) is reproduced as well as the even-odd effect in the $\vert e,N_0,o\rangle$ states (green, \textbf{II.}) in accordance with the experiments. Fig. \ref{multi_island_simulation}\textbf{b} exhibits the qualitative behavior of Fig. \ref{multi_island_measurement_br}\textbf{b}. with the $\vert e,N_0,o\rangle$ (red, \textbf{III.}) and $\vert o,N_0,o\rangle$ (pink, \textbf{IV.}) sectors. The main tendency of $S_{o\mathrm{(L)}}'\lessapprox S_{o\mathrm{(R)}}' <S_o''$ is recovered in the simulations, which is demonstrated in Fig. \ref{multi_island_simulation}\textbf{c}, where we performed the same analysis as for the experimental data in Fig. \ref{multi_island_measurement_br}\textbf{d} with the linecuts taken along the colored arrows of panel \textbf{a-b}. From the spectra, $E_\mathrm{L}=31\,\mu$eV, $E_\mathrm{R}=26\,\mu$eV, and $E_\mathrm{HAM}=19\,\mu$eV binding energies have been estimated with $\Delta S=2(S_o''-S_{o\mathrm{(R)}}')/(S_e+S_o)\approx 9$\% relative spacing reduction. Despite its simplicity and considering only two orbitals for the island, this simple model gives a good agreement with the measurements.

To provide a more realistic description of the SCI, we
also performed DMRG calculations for a superconducting grain with $N=20$ orbitals, spanning a finite bandwidth with level spacing $\delta$. In this more realistic model, levels of the red and green QDs are tunnel-coupled 
to the $i$th orbital in the SCI with amplitudes $t_{\mathrm{L(R)}i}$, as depicted in the inset of Fig. \ref{multi_island_simulation}\textbf{d}. Tuning the individual tunnel couplings allows us to model the mesoscopic randomness of the system, and to define an overlap parameter between the Shiba states, $L=\vert \mathbf{t}_\mathrm{L}\cdot \mathbf{t}_\mathrm{R}\vert / (\vert \mathbf{t}_\mathrm{L}\vert \vert \mathbf{t}_\mathrm{R}\vert )$ with $\mathbf{t}_\mathrm{L(R)}=(t_\mathrm{L(R)1},t_\mathrm{L(R)2},...,t_{\mathrm{L(R)}N}) $. Intuitively, for $L=1$ the coupling is symmetric and all orbitals of the SCI are coupled to the QDs with equal weights, while for $L=0$ the QDs are effectively decoupled. Energies of the ground state and the lowest excited states have been calculated as a function of the QD and SCI electron fillings, $m,n$, and $N_0$, leading to a similar phase diagram as for our simple model. Details of the DMRG computations are in Supplementary Note 3.

Fig. \ref{multi_island_simulation}\textbf{d} shows the DMRG-based excitation energies (solid circles) of the Coulomb-aided YSR states $E_\mathrm{L(R)}$ ($\vert o,o,e\rangle$ and $\vert e,o,o\rangle$ in red and green) and the heteroatomic Andreev molecule state $E_\mathrm{HAM}$ ($\vert o,o,o\rangle$ in pink) as a function of $L$. In the absence of overlap ($L=0$), the heteroatomic state does not gain energy compared to the YSR state residing at lower energy, thus $E_\mathrm{HAM}\approx E_\mathrm{R}$. Increasing $L$ hybridizes the Shiba states, and reduces the energy $E_\mathrm{HAM}$ of the bonding orbital continuously, while $E_\mathrm{L(R)}$ are only slightly affected. This trend confirms that the $\vert o,o,o\rangle$ state is stabilized by the hybridization of the 3 electrons residing on the QDs and on the SCI. In the same panel, the dashed and dotted lines with the corresponding colors display the energies extracted from the experiments and from our simple model, respectively. As one can see, the experimental energies match well the DMRG-based values at $L\approx 0.55$, which indicates that the overlap is significant in our heteroatomic Andreev molecule.

Finally, the DMRG calculation also allows us to reveal SCI-mediated spin correlations between the QDs. 
Fig. \ref{multi_island_simulation}\textbf{e} presents the DMRG-based total spin correlator of the QDs in the ground state, $\langle \mathscr{S}_\mathrm{L}\mathscr{S}_\mathrm{R}\rangle$, versus the electron parity of the green QD, $n$, and the electron number on the SCI, $N_0$, for fixed $m=o$ on the red QD. When the green QD is filled with an even number of electrons ($\vert o,o,e\rangle$, $\vert o,e,e\rangle$ states in green), the correlator ultimately gives 0 due to $\mathscr{S}_\mathrm{R}\approx 0$. In
the $\vert o,e,o\rangle$ state, the correlator takes a finite negative value, reflecting antiferromagnetic ordering, similarly to standard double QD systems. In contrast, for the $\vert o,o,o\rangle$ state, weak ferromagnetic correlations are predicted \cite{bacsi2023exchange}, while a quasiparticle with an antiparallel spin resides on the SCI. This type of coupling can be interpreted as a \emph{superexchange} between the QD spins mediated by the SCI as outlined in Fig. \ref{multi_island_measurement_br}\textbf{c} \textbf{IV}. However, here the superexchange is mediated by the SCI, and the gate control of its parity allows a transition from antiferromagnetic to ferromagnetic exchange. In further experiments, the exchange on the QDs could be studied either by polarizing the spin using an external magnetic field\cite{wang2022singlet} or by micromagnets exploiting the advantage of the large g-factor in the InAs wires\cite{fabian2016magnetoresistance,bordoloi2022spin}.

\section{Conclusions}

In summary, we realized a heteroatomic Andreev molecule as a result of the interplay between an SCI and two QDs.
We performed electrical transport measurements to explore the character of the lowest energy excitations in the SCI. By exploiting the even-odd effect observable in Coluomb blockade spectroscopy of an SCI, we have found that two
electron spins residing in separate QDs can couple to the same quasiparticle at the SCI, and create a pair of hybridized YSR states. We captured the formation of a heteroatomic Andreev molecule from the YSR states by tuning the QDs to the appropriate electron occupations. The experimentally observed signatures have been reproduced by a simple model as well as by more elaborate DMRG-based simulations. The latter also confirmed a significant overlap of YSR states residing in the distinct QDs, and also predicted a ferromagnetic superexchange between the QD spins. The robust hybridization demonstrated in the molecular state is a proof of principle that strong coupling in polyatomic chains can be engineered. Regarding their diversity, SCI-QD hybrids are not only essential towards novel synthetic superconducting 1D crystals but represent also the first steps towards achieving the atomic-level manipulation in SCs\cite{saldana2023richardson} and future polymerization.

\section{Methods}

\textbf{Sample fabrication.} The InAs nanowires were grown by molecular beam epitaxy in the wurtzite phase along the $\langle 0001\rangle$ direction catalyzed by Au. The Au droplets were patterned by electron beam lithography (EBL) which allowed to control the diameter, distance, and the corresponding alignment of the cross-sections of the wires.\cite{kanne2021double} A 20-nm-thick Al layer covering 2 facets was evaporated at low temperature in-situ providing epitaxial, oxide-free layers, which connected the wires. The nanowires were transferred to a p-doped Si wafer capped with 290 nm thick SiO\textsubscript{2} layer by using an optical transfer microscope with micromanipulators. The Al was partially removed by wet chemical etching both on the top and on the bottom leaving a $\approx 700$ nm long SCI in the middle as shown in the scanning electron micrograph (SEM) in Fig. \ref{multi_island_device_and_island}\textbf{a}. The 4 Ti/Au electrodes (yellow) were defined by EBL such that each one contacted only one nanowire segment individually, while 2 wide plunger gates were deposited next to the SCI. In a distinct EBL step, large space-periodic finger gates were installed surrounding the nanowires to control the transport. Electronic transport measurements were performed at a base temperature of $40\,$mK. The voltage on the outer finger gates in Figs. \ref{multi_island_device_and_island}\textbf{a-b} were tuned to form the tunnel barriers of the QDs, while the middle ones were used to tune their level positions. The top left electrode was biased with $V_\mathrm{AC}$ as a source and the rest acted as drains biased with DC voltage $V_\mathrm{SD}$. Differential conductances $G_\mathrm{R}$ and $G_\mathrm{L}$ in the left and right branches were measured simultaneously via the red and green QDs, respectively (the top right arm was floated in these experiments). We note that there was no direct tunneling between the red and green QDs and they were coupled only through the SCI\cite{kurtossy2022parallel}.

\section{Acknowledgments}

The authors are thankful to EK MFA for providing their facilities for sample fabrication. We thank A. P\'alyi, for discussion. This work has received funding Topograph FlagERA, the QuantERA 'SuperTop' (NN 127900) network, the FETOpen AndQC (828948), the COST Nanocohybri network, and from the OTKA K138433 and OTKA FK132146 grants. This research was supported by the Ministry of Innovation and Technology and the NKFIH within the Quantum Information National Laboratory of Hungary and by the Quantum Technology National Excellence Program, UNKP-22-3 and UNKP-23-4 New National Excellence Program of the Ministry for Innovation and Technology, from the source of the National Research, Development and Innovation Fund, Novo Nordisk Foundation project SolidQ, the Carlsberg Foundation, and the Danish National Research Foundation (DNRF 101). 

\section{Author contributions}

O. K. fabricated the device, performed the measurements and did the data analysis. M. B. and Z. S. built the theoretical model and developed the numerical simulations. C. P. M. and Z. G. performed the DMRG calculations. T. K. and J. N. developed the nanowires. All authors discussed the results and worked on the manuscript. P. M. and S. C. proposed the device concept and guided the project.

\section{Competing interests}

The Authors declare no Competing Financial or Non-Financial Interests.

\end{document}


\renewcommand{\theequation}{S\arabic{equation}}
\renewcommand{\figurename}{Supplementary~Figure}
\renewcommand{\tablename}{Supplementary~Table}
\renewcommand{\thesection}{SUPPLEMENTARY~NOTE~\arabic{section}}
\renewcommand{\thetable}{\arabic{table}}

\title{Supplementary Notes for Heteroatomic Andreev molecule in a superconducting island-double quantum dot hybrid}

\author{Oliv\'er K\"urt\"ossy}
\affiliation{Department of Physics, Institute of Physics, Budapest University of Technology and Economics, M\H{u}egyetem rkp. 3, H-1111 Budapest, Hungary}

\author{Mih\'aly Bod\'ocs}
\affiliation{Department of Physics, Institute of Physics, Budapest University of Technology and Economics, M\H{u}egyetem rkp. 3, H-1111 Budapest, Hungary}

\affiliation{MTA-BME Nanoelectronics Momentum Research Group, M\H{u}egyetem rkp. 3, H-1111 Budapest, Hungary}

\author{Catalin Pascu Moca}
\affiliation{HUN-REN—BME Quantum Dynamics and Correlations Research Group,
Budapest University of Technology and Economics, M\H{u}egyetem rkp. 3, H-1111 Budapest, Hungary}

\affiliation{Department of Physics, University of Oradea, 410087, Oradea, Romania}

\author{Zolt\'an Scher\"ubl}
\affiliation{Department of Physics, Institute of Physics, Budapest University of Technology and Economics, M\H{u}egyetem rkp. 3, H-1111 Budapest, Hungary}

\affiliation{MTA-BME Nanoelectronics Momentum Research Group, M\H{u}egyetem rkp. 3, H-1111 Budapest, Hungary}

\author{Ella Nikodem}
\affiliation{Physics Institute II, University of Cologne, Zülpicher Str. 77, 50937 Cologne, Germany}

\author{Thomas Kanne}
\affiliation{Center for Quantum Devices, Niels Bohr Institute, University of Copenhagen, 2100 Copenhagen, Denmark}

\author{Jesper Nyg{\aa}rd}
\affiliation{Center for Quantum Devices, Niels Bohr Institute, University of  Copenhagen, 2100 Copenhagen, Denmark}

\author{Gergely Zar\'and}
\affiliation{HUN-REN—BME Quantum Dynamics and Correlations Research Group,
Budapest University of Technology and Economics, M\H{u}egyetem rkp. 3, H-1111 Budapest, Hungary}

\affiliation{Department of Theoretical Physics, Institute of Physics,
Budapest University of Technology and Economics, M\H{u}egyetem rkp. 3, H-1111 Budapest, Hungary}

\author{P\'eter Makk}
\email{makk.peter@ttk.bme.hu}
\affiliation{Department of Physics, Institute of Physics, Budapest University of Technology and Economics, M\H{u}egyetem rkp. 3, H-1111 Budapest, Hungary}

\affiliation{MTA-BME Correlated van der Waals Structures Momentum Research Group, M\H{u}egyetem rkp. 3, H-1111 Budapest, Hungary}

\author{Szabolcs Csonka}
\email{csonka.szabolcs@ttk.bme.hu}
\affiliation{Department of Physics, Institute of Physics, Budapest University of Technology and Economics, M\H{u}egyetem rkp. 3, H-1111 Budapest, Hungary}

\affiliation{MTA-BME Nanoelectronics Momentum Research Group, M\H{u}egyetem rkp. 3, H-1111 Budapest, Hungary}

\date{\today}
\maketitle

\section{Coulomb blockade spectroscopy}

\begin{figure}[h!]
\begin{center}
\includegraphics[width=1\textwidth]{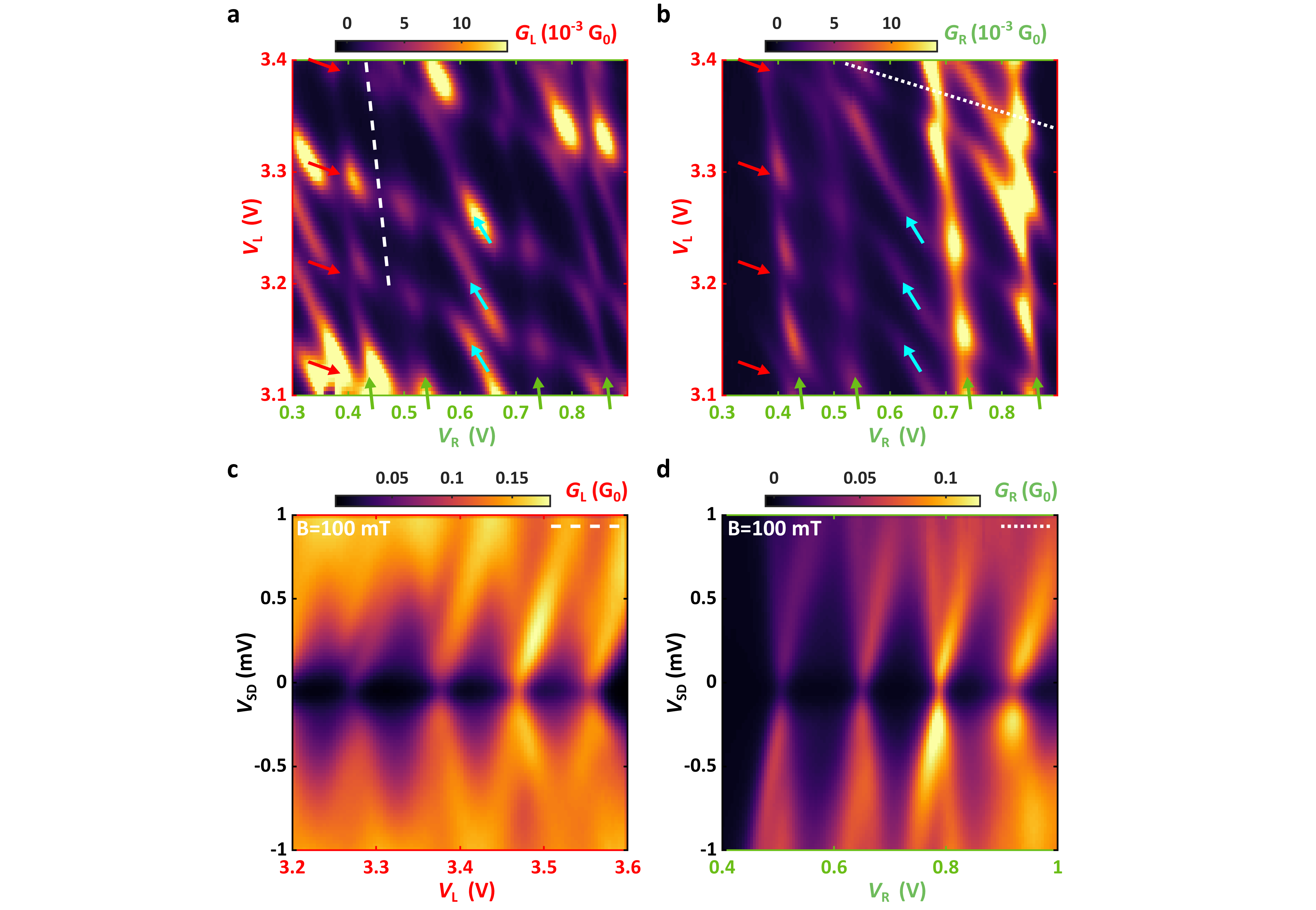}
\caption{\label{multi_island_diamonds} \textbf{Characterization of the QDs.} \textbf{a} $G_\mathrm{L}$ and \textbf{b} $G_\mathrm{L}$ versus $V_\mathrm{L}$ and $V_\mathrm{R}$ plunger gate voltages in the superconducting states. Resonances with 3 different slopes are present: the ones indicated by the red, green, and cyan arrows belong to the red, green QD, and the SCI, respectively. \textbf{c-d} Finite-bias spectroscopy along the white dashed and dotted line from panels \textbf{a-b} in the normal state. The charging energies of the QDs exceed the SCI's one, while there is a finite level spacing on them as well.}
\end{center}
\end{figure}

Supp. Figs. \ref{multi_island_diamonds}\textbf{a-b} show the zero-bias $G_\mathrm{L}$ and $G_\mathrm{R}$ measured via the red and green QDs as a function of their plunger gate voltages, $V_\mathrm{L}$ and $V_\mathrm{R}$. In the stability maps, there are several resonant lines with 3 dominant lever arms. The one indicated by the red arrows belongs to the red QD, which is mostly visible in panel \textbf{a}, where it is the local signal. The resonances marked by the green arrows are attributed to the green QD, therefore they dominate in panel \textbf{b}. The diagonal lines with the cyan arrows are the SCI resonances, which appear in both diagrams as its signal is measured in $G_\mathrm{L}$ and $G_\mathrm{R}$ as well. Finite-bias spectroscopies were performed along the white dashed and dotted lines in the normal state (achieved by a $B=100\,$mT out-of-plain magnetic field) revealing the Coulomb diamonds of the red and green QDs in Supp. Figs. \ref{multi_island_diamonds}\textbf{c} and \ref{multi_island_diamonds}\textbf{d}, respectively. From the size of the diamonds in the examined range, charging energies and level spacings $U_\mathrm{L}\approx 0.4\,$meV and $\delta_\mathrm{L}\approx 0.1\,$meV were found for the red QD, while $U_\mathrm{R}\approx 1.2\,$meV and $\delta_\mathrm{R}\approx 0.35\,$meV were derived for the green one.

In Figs. 3\textbf{a-b} of the main text, the stability map of the SCI and the green QD was explored while the on-site energy of the red QD was fixed. To maintain the level position in the red, untuned QD, the gate voltage on its plunger gate was compensated while the other one was ramped. The size of the voltage compensation was defined by the lever arm ratio referring to the cross capacitance strength of the red and green QDs and their gate electrodes, which was $\alpha_\mathrm{L,R}\approx 1/6$. This quantity was $\alpha_\mathrm{L(R),SCI}\approx 1/10$ for the red (green) QD and the SCI. Since $V_\mathrm{SC}$ was varied in a small window compared to $V_\mathrm{L(R)}$, its gating effect imposed on the untuned red (green) QD was neglected.

\section{Additional data}

\begin{figure}[h!]
\begin{center}
\includegraphics[width=1\textwidth]{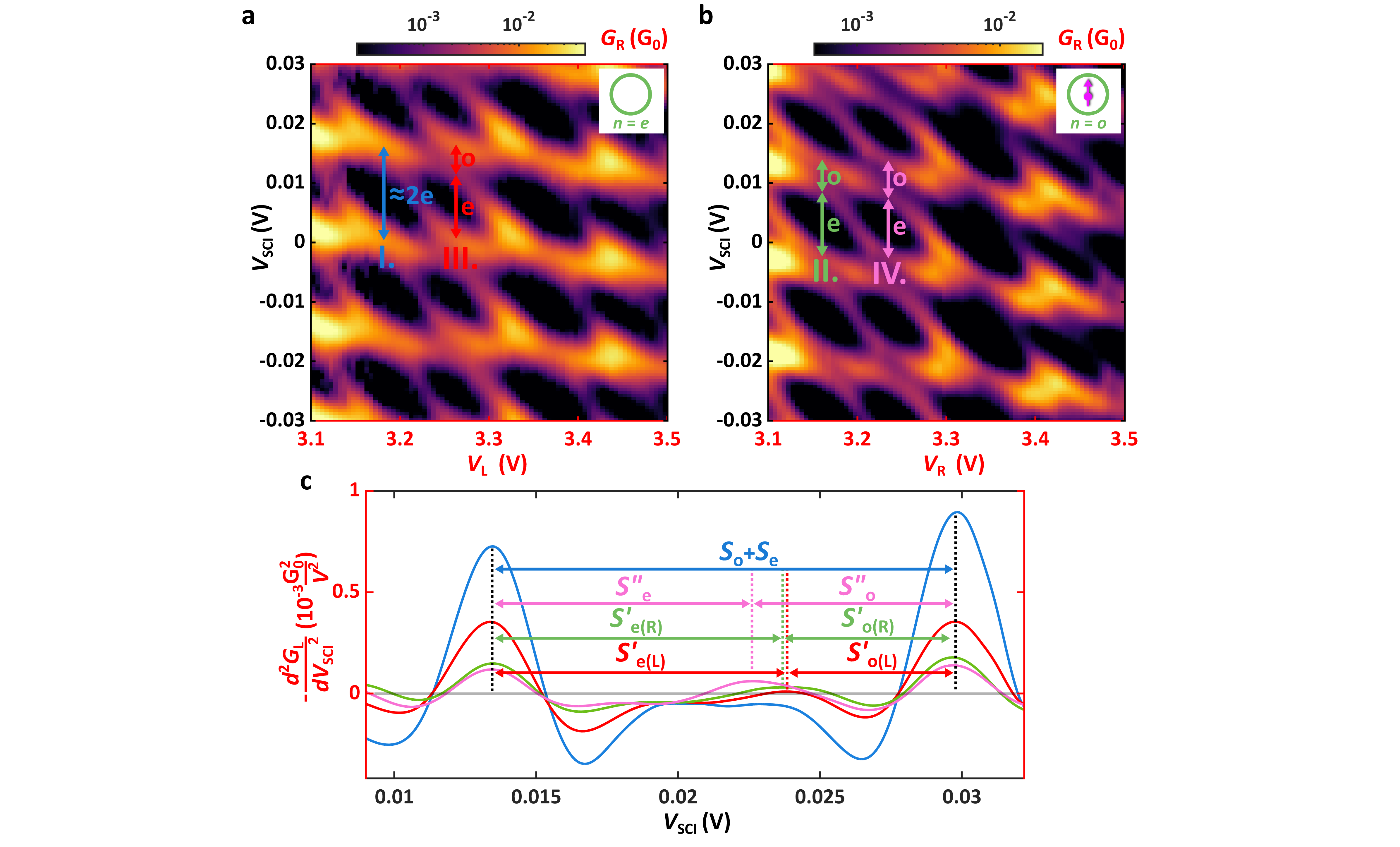}
\caption{\label{multi_island_measurement_bl} \textbf{Stability diagrams at different QD occupations (with swapped red and green QD roles).} \textbf{a} Zero-bias stability sweep as a function of $V_\mathrm{L}$ and $V_\mathrm{SCI}$ via the SCI-red double QD with even number of electrons in the green QD. The $\vert e,N_0,e\rangle$ diamonds (\textbf{I.)} at ($V_\mathrm{L}=3.2\,$V) reflect 2e charging, for $\vert o,N_0,e\rangle$ (at $V_\mathrm{L}=3.27\,$V, \textbf{III.}), the even-odd effect is recovered. \textbf{b} Same as \textbf{a}, but with $\vert m,N_0,o\rangle$ configurations examined. The odd sector of the SCI, $\vert o,N_0,o\rangle$, is broadened further from $S'_{o\mathrm{(L)}}$ (III.) to $S''_o$ (IV.) revealed by the pink arrows. This suggests the hybridization of the YSR states captured now in both $G_\mathrm{L}$ and $G_ \mathrm{R}$. \textbf{d} Peak analysis of SCI resonance lines taken along the colored arrows in panels \textbf{a-b}. $S'_{o\mathrm{(L)}}\lessapprox S'_{o\mathrm{(R)}} <S''_o$ fulfills the expectation predicting a polyatomic Andreev molecule. }
\end{center}
\end{figure}

In the stability maps, qualitatively the same behavior was obtained, when the red QD was tuned together with the SCI and the green QD occupation was fixed instead (oppositely to Fig. 3 in the main text). Supp. Fig. \ref{multi_island_measurement_bl}\textbf{a} demonstrates $G_\mathrm{L}$ at zero bias as a function of $V_\mathrm{L}$ and $V_\mathrm{SCI}$ with even number of electrons in the green QD (as shown by the inset) allowing to explore the even-odd effect of the $\vert m,N_0,e\rangle$ states. The vertical pattern at $V_\mathrm{L}=3.2\,$V exhibits the close 2e charging of the $\vert e,N_0,e\rangle$ states (\textbf{I.}) as expected from the previous results of Fig. 3\textbf{a} from the main text. At $V_\mathrm{L}=3.27\,$V, one can see the splitting of the SCI resonances yielding a finite even-odd effect in the $\vert o,N_0,e\rangle$ configurations (\textbf{III.}, see the red arrows), which corresponds to the signature of the Coulomb-aided YSR singlet living in the red QD. Supp. Fig. \ref{multi_island_measurement_bl}\textbf{b} shows the same map as panel \textbf{a} with the difference of having a single electron in the green QD, thereby mapping the $\vert m,N_0,o\rangle$ sectors. Here the close 2e charging of case \textbf{I.} from panel \textbf{a} turns into a single one of case \textbf{II.} ($\vert e,N_0,o\rangle$) marked by the green arrows, which is the evidence of a YSR state formed in the green QD captured now in the signal of the bottom left arm. The $\vert o,N_0,o\rangle$ configurations (\textbf{IV.}) have the widest odd sector gain $S'_{o\mathrm{(L)}}\lessapprox S'_{o\mathrm{(R)}} <S''_o$ as depicted by the pink arrows. These observations concluded from Figs. \ref{multi_island_measurement_bl}\textbf{a-b} are consistent with the ones in Figs. 3\textbf{a-b} in the main text.

The analysis introduced in Fig. 3\textbf{d} in the main text gives a similar outcome when it is applied to the data of $G_\mathrm{L}$ in Supp. Figs. \ref{multi_island_measurement_bl}\textbf{a-b}, which is shown in Fig. \ref{multi_island_measurement_bl}\textbf{c} for the completeness. The calculated $q=-d^2G_\mathrm{L}/dV_\mathrm{SCI}^2$ curves are taken along the colored arrows, and although the secondary peaks are rather small, $q\geq 0$ and $S'_{o\mathrm{(L)}}\lessapprox S'_{o\mathrm{(R)}} <S''_o$ still hold. One can see that the highlighted odd spacings $S''_o$, $S'_{o\mathrm{(R)}}$, $S'_{o\mathrm{(L)}}$ of the pink, green, and red signals are in good agreement with the ones in Fig. 3\textbf{d} from the main text, and we estimate $E_\mathrm{L(R)}\approx 32\, \mu$eV and $E_\mathrm{HAM}\approx 16\, \mu$eV from this set of data. This series of measurements strengthens our hypothesis about the polyatomic Andreev molecule spatially extending over the QDs and the SCI since the modulation of the even-odd effect was captured in both $G_\mathrm{L}$ and $G_\mathrm{R}$.

\section{Modeling}
\subsection{Mixed orbital Anderson model}

\begin{figure}[h!]
\begin{center}
\includegraphics[width=0.5\textwidth]{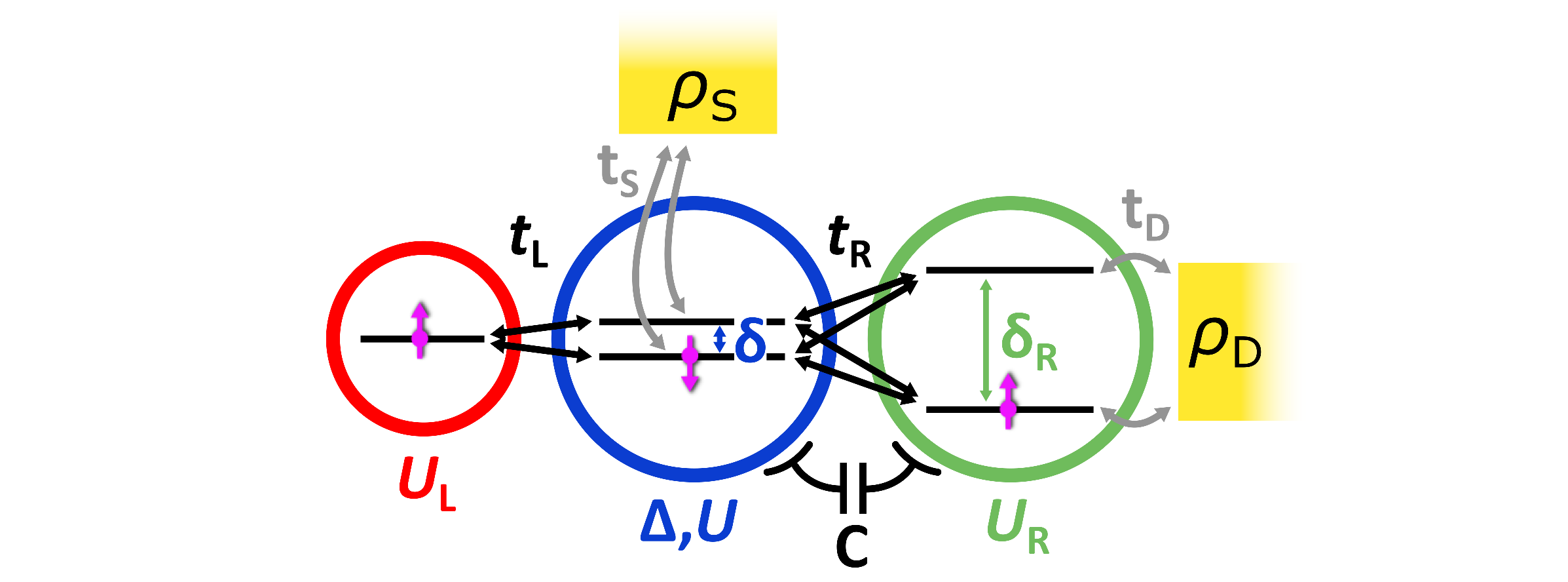}
\caption{\label{multi_island_model} \textbf{Schematics of the polyatomic Andreev molecule model.} The SCI and the green QD are modeled by 2-2 orbitals, while the red QD is by a single one. All tunnel couplings are shown by the arrows between the levels. The SCI and the green QDs are connected to the source and drain leads with a constant density of states $\rho_\mathrm{S}$, $\rho_\mathrm{D}$ (yellow rectangles).}
\end{center}
\end{figure}

In this section, we introduce the minimal model we developed, which can reproduce the main experimental findings. The SCI is represented using a Richardson model with only two levels, supplemented with ("R") QD is modeled by a two-level Anderson model, while the red ("L") QD is represented by a single-level Anderson model. Consistent with the experimental setup, the topology is as shown in Supp. Fig. \ref{multi_island_model}, where tunneling occurs between the red QD-SCI, and between the SCI-green QD. No direct tunneling is present between the QDs. Additionally, we account for the capacitive coupling of the SCI and the green QD.

The schematic of the model is illustrated in Fig. \ref{multi_island_model}. The SCI consists of $N=2$ orbitals in the Richardson picture with a common charging energy $U$ and superconducting gap $\Delta$. One of the orbital energy $\epsilon_\mathrm{1} $ is chosen to be 0, thus the level spacing is set small $\delta=\epsilon_\mathrm{2}=U/100$ typical for metallic islands. The green QD is treated in the 2-orbital Anderson model with charging energy $U_\mathrm{R}$ and level spacing $\delta_\mathrm{R}=\epsilon_\mathrm{R2}$ ($\epsilon_\mathrm{R1}=0$ applies here as well). The red QD is handled as a single level to keep the model minimal. The SCI is tunnel coupled to both levels of the green QD and the single level of the red QD, however, the QDs are not connected directly as shown in Fig. \ref{multi_island_model}. We also consider the mutual capacitance $C$ between the SCI and the green QD, but the cross capacitances to the red QD are neglected. The total Fock-space Hamiltonian of the system is composed as
\begin{equation}\label{eq:shiba_molecule}
H_\mathrm{HAM}=H_\mathrm{SCI}+H_\mathrm{R}+H_\mathrm{L}+H_\mathrm{TR}+H_\mathrm{TL},
\end{equation}
where
\begin{equation}\label{eq:shiba_molecule_detailed}
\begin{aligned}
H_\mathrm{SCI}&= \left(\varepsilon+\frac{\Delta}{2}\right)\sum_i n_{i} +\left(\sum_i n_{i}\right)\left(\sum_j n_{j}-1\right)\frac{U}{2}+\sum_i n_{i}\epsilon_{i}\\&-\frac{\Delta}{N}\underbrace{\sum_{i,j}c^\dagger_{i\uparrow}c^\dagger_{i\downarrow}c_{j\downarrow}c_{j\uparrow}}_{W}  \\
H_\mathrm{R}&= \varepsilon_\mathrm{R}\sum_\alpha n_{\mathrm{R}\alpha} +\left(\sum_\alpha n_{\mathrm{R}\alpha}\right)\left(\sum_\beta n_{\mathrm{R}\beta}-1\right)\frac{U_\mathrm{R}}{2}+\sum_\alpha n_{\mathrm{R}\alpha}\epsilon_{\mathrm{R}\alpha}  \\
H_\mathrm{L}&=\varepsilon_\mathrm{L} n_{\mathrm{L}} +\frac{U_\mathrm{L}}{2} n_\mathrm{L}\left( n_\mathrm{L}-1\right)  \\
H_\mathrm{TR}&=t_\mathrm{R} \left(\sum_{\alpha i\sigma} d_\mathrm{R\alpha\sigma}^{\dagger}c_{i\sigma}+ c_{i\sigma}^{\dagger}d_\mathrm{R\alpha\sigma}\right) + C\left(\sum_i n_{i}\right)\left(\sum_\alpha n_{\mathrm{R}\alpha}\right) \\
H_\mathrm{TL}&=t_\mathrm{L} \left(\sum_{i\sigma} d_\mathrm{L\sigma}^{\dagger}c_{i\sigma}+ c_{i\sigma}^{\dagger}d_\mathrm{L\sigma}\right) .
\end{aligned}
\end{equation}

In the equations above, $n_{\mathrm{R}\alpha\mathrm{(L)}}$ is the particle number operator of orbital $\alpha$ in the green (red) QD with $d^{\left(\dagger\right)}_{\mathrm{R}\alpha \mathrm{[BL]}\sigma}$ being the annihilation (creation) operator of an electron with spin $\sigma$, while $\varepsilon_\mathrm{R(L)}$ is the green [red] QD on-site energy. In addition, $n_{i}$ is the particle number operator of orbital $i$ in the SCI with $c^{\left(\dagger\right)}_{i\sigma}$ being the annihilation (creation) operator of an electron with spin $\sigma$, and $\varepsilon$ is SCI on-site energy. We used the notations of $n(m)=\langle n_\mathrm{R(L)}\rangle$ and $N_0=\langle n\rangle$. $t_\mathrm{R(L)}$ is the hopping amplitude between one of the levels of the SCI and the green (red) QD. 

We briefly comment on the combinatorial source of the $\Delta /N$ normalization prefactor there. According to Eq. 1 in the main text, the even-odd effect vanishes, when $\Delta =U$ is satisfied. Even so, the $N_0=0$, $N_0=1$, and $N_0=2$ states are degenerate, which sets the
\begin{equation}\label{eq:w_condition}
A\langle W\rangle=U\Big|_{\Delta=U}
\end{equation}
condition for 2e charging, where $A$ is the desired normalization constant. Assuming $\epsilon_i\approx\epsilon_j=\epsilon$ (negligible level spacing on the SCI), the orbitals are filled with $1/\sqrt{N}$ amplitude in the ground state. Therefore the diagonal contribution of $\langle W\rangle$ is $1$ (since the 2 electrons can be annihilated and created on $N$ orbitals), while the off-diagonal one is $N-1$ (as the 2 electrons can be scattered from $N$ orbitals to $N-1$ new ones). Substituting the $\langle W\rangle =N$ result into Eq. \ref{eq:w_condition}, $A=\Delta /N$ is given.

The parameters included in the model are summarized in Supp. Table 1.

\begin{table}[H]
\centering
\begin{tabular}{|c|c|c|c|c|c|c|c|c|}
\hline 
Parameter & $U_\mathrm{L}$ & $U_\mathrm{R}$ & $U$ & $\Delta$ & $C$ & $\epsilon_\mathrm{R}$ & $t_\mathrm{L}$ & $t_\mathrm{L}$ \\ 
\hline 
meV & 0.4 & 1.2 & 0.085 & 0.075 & 0.04 & 0.35 & 0.035 & 0.06 \\ 
\hline 
\end{tabular} 
\caption{\label{parameter_table}}
\end{table}

Direct diagonalization was performed on $H_\mathrm{HAM}$ to derive the ground state wave functions and the electron occupations as a function of $\varepsilon, \varepsilon_\mathrm{R}$, and $\varepsilon_\mathrm{L}$. Transport through the green QD-SCI double QD (while coupled to the red one as well) was calculated by solving the Master equation in the stationary limit:
\begin{equation}\label{eq:shiba_molecule_master}
\frac{dP_{\chi}}{dt} = \sum_{\chi'\neq\chi} \left( W_{\chi\chi'} P_{\chi'} - W_{\chi'\chi} P_{\chi}\right).
\end{equation}
$P_{\chi}$ is the occupation probability of the eigenstate $\chi$ with the constriction of $\sum_{\chi}P_{\chi}=1$.
$W_{\chi\chi'}$ denotes the total transition rate from $\ket{\chi'}$ state to $\ket{\chi}$, which is calculated by Fermi's golden rule. $W_{\chi\chi'}$ is the sum of
\begin{equation}\label{eq:shiba_molecule_rates}
\begin{aligned}
W_{\chi'\chi}\left( c^{\dagger}_{i\sigma} \right) &= \Gamma_\mathrm{S}\left| \bra{\chi'} c^{\dagger}_{i\sigma}  \ket{\chi} \right|^2 \ f\left(E_{\chi}-E_{\chi'}-eV_\mathrm{AC} \right) \\
W_{\chi'\chi}\left( c_{i\sigma} \right) &= \Gamma_\mathrm{S}\left| \bra{\chi'} c_{i\sigma}  \ket{\chi} \right|^2  \left( 1-f\left(E_{\chi'}-E_{\chi}-eV_\mathrm{AC} \right)\right) \\
W_{\chi'\chi}\left( d^{\dagger}_{\mathrm{R}\alpha\sigma} \right) &= \Gamma_\mathrm{D}\left| \bra{\chi'} d^{\dagger}_{\mathrm{R}\alpha\sigma}  \ket{\chi} \right|^2  f\left(E_{\chi}-E_{\chi'}+eV_\mathrm{AC} \right) \\
W_{\chi'\chi}\left( d_{\mathrm{R}\alpha\sigma} \right) &= \Gamma_\mathrm{D}\left| \bra{\chi'} d_{\mathrm{R}\alpha\sigma}  \ket{\chi} \right|^2  \left( 1- f\left(E_{\chi'}-E_{\chi}+eV_\mathrm{AC} \right)\right).\\
\end{aligned}
\end{equation}
$\Gamma_{\mathrm{S(D)}}=\pi t_{\mathrm{S(D)}}^2\rho_{\mathrm{S(D)}}(0)$ is the coupling strength of the SCI (green QD) to one of the normal leads with density of states $\rho_{\mathrm{S(D)}}(0)=\,$const. as depicted in Fig. \ref{multi_island_model}.
The stationary current is given by solving Eq. \ref{eq:shiba_molecule_master} at $dP_{\chi}/dt=0$. We calculate the differential conductance at the green QD-normal lead interface as
\begin{equation}\label{eq:shiba_molecule_diff_g}
G_\mathrm{R} = \frac{dI_\mathrm{R}}{dV_\mathrm{AC}}=\frac{e}{\hbar V_\mathrm{AC}} \sum_{\alpha\chi\chi'\sigma} \left( W_{\chi'\chi}\left(d_{\mathrm{R}\alpha\sigma} \right) - W_{\chi'\chi}\left(d^{\dagger}_{\mathrm{R}\alpha\sigma} \right) \right) P_{\chi}.
\end{equation}

\subsection{Richardson model}

To achieve a more accurate calculation, we present a detailed description of the superconducting island side-coupled to two QDs. The Hamiltonian of the device is given by:
\begin{equation}
    H_{\rm device} = H_{\rm SCI} + H_{\rm QDs} + H_{\rm tunneling} \label{eq:full_H}
\end{equation}

We model the superconducting island using the Richardson model\cite{RICHARDSON.1963,vonDelft.2001}:
\begin{equation}
    H = \sum_{i\sigma} \epsilon_i c_{i\sigma}^\dagger c_{i\sigma} - \alpha d \sum_{ij} c_{i\uparrow}^\dagger c_{i\downarrow}^\dagger c_{j\downarrow} c_{j\uparrow} + E_C \left(\hat{N} - N_0\right)^2 \label{eq:Richardson}
\end{equation}
where $\epsilon_i$ are the discrete single-particle energy levels, $c_{i\sigma}^{(\dagger)}$ are the annihilation (creation) operators for electrons with spin $\sigma$ in state $i$, $\alpha$ is the parity interaction strength of the contact interaction, and $\hat{N} = \sum_{i} c_{i\sigma}^\dagger c_{i\sigma}$ represents the particle number in the superconducting island (SCI). In our convention, $d$ stands for the level spacing while the single-particle energies are equidistant levels $\epsilon_i = -D + (i - 1/2)d - \alpha d/2$, with $2D$ being the energy bandwidth. By subtracting the last term in the energy spectrum, the particle-hole symmetry is recovered\cite{Pavesic.2021}. In the weak coupling regime $\alpha \ll 1$, the superconducting gap can be recovered as\cite{vonDelft.2001}:
\begin{equation}
    \Delta \approx 2D\, e^{-1/\alpha}
\end{equation}

In Eq.~\eqref{eq:Richardson}, $E_C$ represents the charging energy of the SCI. In our numerical simulations, we set $D = 1$ as the energy unit. The number of levels inside the superconductor was fixed at 20, resulting in a level spacing of $d = 0.1$. Additionally, we set $\alpha = 0.95$, which yielded a superconducting gap $\Delta = 0.75$. To model our experimental data, the charging energy $E_C$ was fixed at $E_C = 0.85$.
When compared to the experimental value of $\Delta = 75 \, \mu\text{eV}$, the energy unit is determined to be $D = 100 \, \mu\text{eV}$.

The Hamiltonian describing the two QDs is:
\begin{equation}
    H_{\rm QDs} = \frac{1}{2} \sum_{a = \{\mathrm{L, R}\}} U_a (n_a - \nu_a)^2
\end{equation}
with $U_a$ being the Coulomb energies in the two QDs, $\nu_a$ controlling the filling of each QD, and $n_a  = \sum_{\sigma = \{\uparrow, \downarrow\}} d_{a\sigma}^\dagger d_{a\sigma}$ representing the particle number operator in QD $a$. Here, $d_{a\sigma}^{(\dagger)}$ are the annihilation (creation) operators for electrons in each QD. In units of $D$, 
the Coulomb energies are fixed to $U_\mathrm{L}= 4$ and $U_\mathrm{R}=12$ respectivly. 
The last term in Eq.~\eqref{eq:full_H} describes the tunneling Hamiltonian between the SCI and the two QDs:
\begin{equation}
    H_{\rm tunneling} = \sum_{a = \{\mathrm{L, R}\}} \sum_{i\sigma} t_{ai} \left( c_{i\sigma}^\dagger d_{a\sigma} + d_{a\sigma}^\dagger c_{i\sigma} \right)
\end{equation}
In our convention, each QD is randomly coupled to all the levels in the SCI. The set of hopping parameters can be described by the vectors $\mathbf{t}_a = (t_{a1}, t_{a2}, \ldots)$, which are picked from a random normal distribution with zero mean and standard deviation $\sigma = d$. 

The strengths of the couplings to each QD are given by $\Gamma_a = \pi \nu_0 |\mathbf{t}_a|$, where $\nu_0$ is the density of states of the SCI assumed in the normal state. Specifically, they are fixed to $\Gamma_\mathrm{L} = 1.13$ and $\Gamma_\mathrm{R}=2.64$ in units of $D$. 
In our numerics, we generate the random coupling for a fixed overlap $L=\vert \mathbf{t}_\mathrm{L}\cdot \mathbf{t}_\mathrm{R}\vert / (\vert \mathbf{t}_\mathrm{L}\vert \vert \mathbf{t}_\mathrm{R}\vert )$. When $L$ is set to 1, the two QDs are coupled with the same amplitudes to each of the levels of the SCI, while in the opposite limit $L = 0$, the two vectors $\mathbf{t}_L$ and $\mathbf{t}_R$ are orthogonal, and the two QDs are completely decoupled.

The Richardson model with finite charging energy, when coupled to superconducting QDs, cannot be addressed using the standard numerical renormalization group approach\cite{Wilson.1975} typically employed in quantum impurity problems. This limitation arises because the superconducting bath is interacting\cite{Pavesic.2021}. 
In order to overcome this limitation, we employ the DMRG method\cite{White.1992,Schollwock.20005} using the matrix product states (MPS) formalism\cite{SCHOLLWOCK.2011} as implemented in the iTensor library\cite{itensor}. This approach is well-suited for handling problems where the Hamiltonian exhibits long-range and all-to-all interactions\cite{Pavesic.2021}.

To determine the ground state energies, average occupations of the QDs, and spin-spin correlations, we conduct extensive calculations within the parameter space $(\nu_L, N_0, \nu_R)$, while keeping the total number of electrons in the system fixed as either even or odd.